\documentclass[aps,prx,longbibliography,twocolumn,notitlepage,citeautoscript,aps,multicol,amsmath,amssymb,longbibliography]{revtex4-1}

\usepackage{graphicx}
\usepackage{epstopdf}
\usepackage{bm}
\usepackage{subfigure}
\usepackage{sidecap}
\graphicspath{{Figures/}}

\newcommand{\si}{\sigma}
\newcommand{\al}{\alpha}
\newcommand{\bt}{\beta}

\newcommand{\vare}{\varepsilon}

\newcommand{\De}{\Delta}

\newcommand{\bl}{\begin{aligned}}
\newcommand{\el}{\end{aligned}}

\newcommand{\be}{\begin{equation}}
\newcommand{\ee}{\end{equation}}

\newcommand{\bea}{\begin{eqnarray}}
\newcommand{\eea}{\end{eqnarray}}
\newcommand{\bd}{\begin{displaymath}}
\newcommand{\ed}{\end{displaymath}}
\newcommand{\ba}{\begin{array}}
\newcommand{\ea}{\end{array}}
\newcommand{\bi}{\begin{itemize}}
\newcommand{\ei}{\end{itemize}}
\newcommand{\bc}{\begin{center}}
\newcommand{\ec}{\end{center}}
\newcommand{\bfl}{\begin{flushleft}}
\newcommand{\efl}{\end{flushleft}}
\newcommand{\bfr}{\begin{flushright}}
\newcommand{\efr}{\end{flushright}}

\newcommand{\hh}{\hat{h}}

\newcommand{\hG}{\hat{G}}

\newcommand{\hbz}{\hat{{\bf z}}}
\newcommand{\hbx}{\hat{{\bf x}}}

\newcommand{\ua}{\uparrow}
\newcommand{\da}{\downarrow}

\newcommand{\tq}{\tilde{q}}
\newcommand{\tbq}{\tilde{{\bf q}}}

\newcommand{\tC}{\tilde{C}}

\newcommand{\om}{i\omega_n}
\newcommand{\omp}{i\omega_{n'}}

\newcommand{\num}{i\nu_m}

\newcommand{\fs}{\frac{1}{2}}

\def\ket#1{\left\vert #1 \right\rangle}
\def\dg{^{\dagger}}

\def\bk{{\bf k}} \def\bq{{\bf q}} 
  \def\bb{{\bf b}}
\def\bQ{{\bf Q}} 
  
 \def\bd{{\bf d}}  
  \def\hbz{\hat{{\bf z}}}
\def\hbx{\hat{{\bf x}}}  

\def\da{\downarrow} \def\ua{\uparrow}
 
\def\dg{\dagger}

\def\bra{\langle}
\def\ket{\rangle}
\def\={\!\!\!&=&\!\!\!}
\def\+{\!\!\!&&\!\!\!+~}
\def\-{\!\!\!&&\!\!\!-~}

\usepackage[colorlinks=true,citecolor=blue]{hyperref}

\usepackage{hyperref,cleveref}

\begin{document}
\date{\today}
\title{Dynamical magnetic response in superconductors with finite momentum pairs}

\author{Peter Thalmeier and Alireza Akbari}
\affiliation{Max Planck Institute for the  Chemical Physics of Solids, D-01187 Dresden, Germany}

\begin{abstract}
We derive the dynamical magnetic response functions in the Fulde-Ferrell (FF) state of a superconductor with inversion symmetry.
The pair momentum $2q$ is obtained by minimization of the condensation energy and the resulting quasiparticle states and spectral functions exhibit the segmentation into paired and unpaired regions due to the finite $q$. The dynamical magnetic susceptibility is then calculated in linear response formalism in the FF state with
finite-q condensate resulting from s-wave or d-wave pairing. 
We show that quasiparticle excitations inside as well as between paired and unpaired segments contribute to the dynamical response. We discuss its dependence on  frequency and  momentum transfer which develops a characteristic symmetry-breaking parallel to q. Furthermore we investigate the possible influence on Knight shift and in the case of d-wave pairing on the spin resonance formation in the FF state.
\end{abstract}


\maketitle

\section{Introduction}
\label{sec:introduction}

In singlet superconductors with modest orbital pair breaking a state with finite-momentum may become stable
at low temperature and high fields. In this state Cooper pairs $(-\bk+\bq\ua,\bk+\bq\da)$ with finite common momentum 
$2\bq$ form the Fulde-Ferrell (FF) \cite{fulde:64} phase. A related phase is the Larkin-Ovchinnikov (LO) \cite{larkin:65} state with superposition of pairs having  (\bq,-\bq) momenta, only the former state will be considered here. These phases have been investigated in detail by various theoretical techniques, mostly focused on the B-T phase diagram and its critical curves. Superconductors of different dimensionality  \cite{machida:84,shimahara:94,shimahara:98,vorontsov:05,mizushima:14} as well as condensed quantum gases \cite{sheehy:07,sheehy:15}. have been studied.\\

Experimental evidence for this exotic pair state in the low-temperature and high-field sector is, however, hard to obtain, possibly caused by sensitivity to impurity scattering \cite{takada:70,matsuda:07,wang:07} and orbital pair breaking \cite{gruenberg:66,adachi:03}. Candidates are found among unconventional heavy fermion superconductors \cite{matsuda:07}, organic materials \cite{lortz:07,mayaffre:14} and also Fe-pnictides \cite{burger:13,zocco:13,matsuda:07}. The evidence for the FF or LO phases is primarily obtained from thermodynamic anomalies  \cite{bianchi:03} or NMR experiments  \cite{kumagai:11} and these results can be used to determine the FFLO phase boundaries.\\

Such experiments, however, do not probe the microscopic nature of the FF state whose central aspect is the breakup of the Fermi surface into paired and unpaired segments. This state is due to a \bk- dependent tradeoff between the loss of condensation energy due to the pair kinetic energy associated with the  overall momentum and a gain in Zeeman energy due to population imbalance of spin states in the external field \cite{combescot:07,zwicknagl:11}.
The relative size of paired and unpaired segments depends on the size of the field where the former vanishes above the critical  field. Probing the microscopic structure of the FF state in practice has rarely been attempted due to lack of suitable techniques. It was proposed \cite{akbari:16,akbari:22} that STM-based quasiparticle interference method is a promising candidate for this purpose.\\

Another important probe for the FF state may be inelastic neutron scattering (INS) which probes the dynamical spin susceptibility. The latter is determined by the quasiparticle excitations in the FF phase which are considerably different from the BCS phase for two reasons: i) the gap amplitude will be much reduced for states in the paired segments and ii) the appearance of unpaired states will lead to additional low energy response and a symmetry breaking in momentum space with respect to the direction of the  pair momentum $2\bq$. Both effects should leave their signature on the dynamical spin response observable by INS. Spin dynamics has sofar been mostly investigated for the $\bq=0$ BCS phase. It also encompasses the possibility of a spin exciton or resonance inside the gap for a unconventional, e.g. d-wave gap symmetry with nodal structure \cite{thalmeier:16,eschrig:06} if quasiparticle exchange interactions are sufficiently strong. Furthermore the static or low energy spin response determines the Knight shift and NMR relaxation rate which is also an important means to investigate the superconducting gap function. For the application of these methods to the FF phase it is therefore necessary to have a detailed theory of the static and dynamical magnetic susceptiblility in this exotic state available for comparison. In the present work we give a derivation of the magnetic response function and a discussion of possible observable principal features. This type of spectroscopic knowledge may contribute to the  microscopic understanding of the  peculiar superconducting  states with finite- momentum Cooper pairs.\\

\section{Model ground state energy of the  FFLO superconductor}
\label{sec: FFLO}

\begin{figure}[t]
\includegraphics[width=0.99\linewidth]{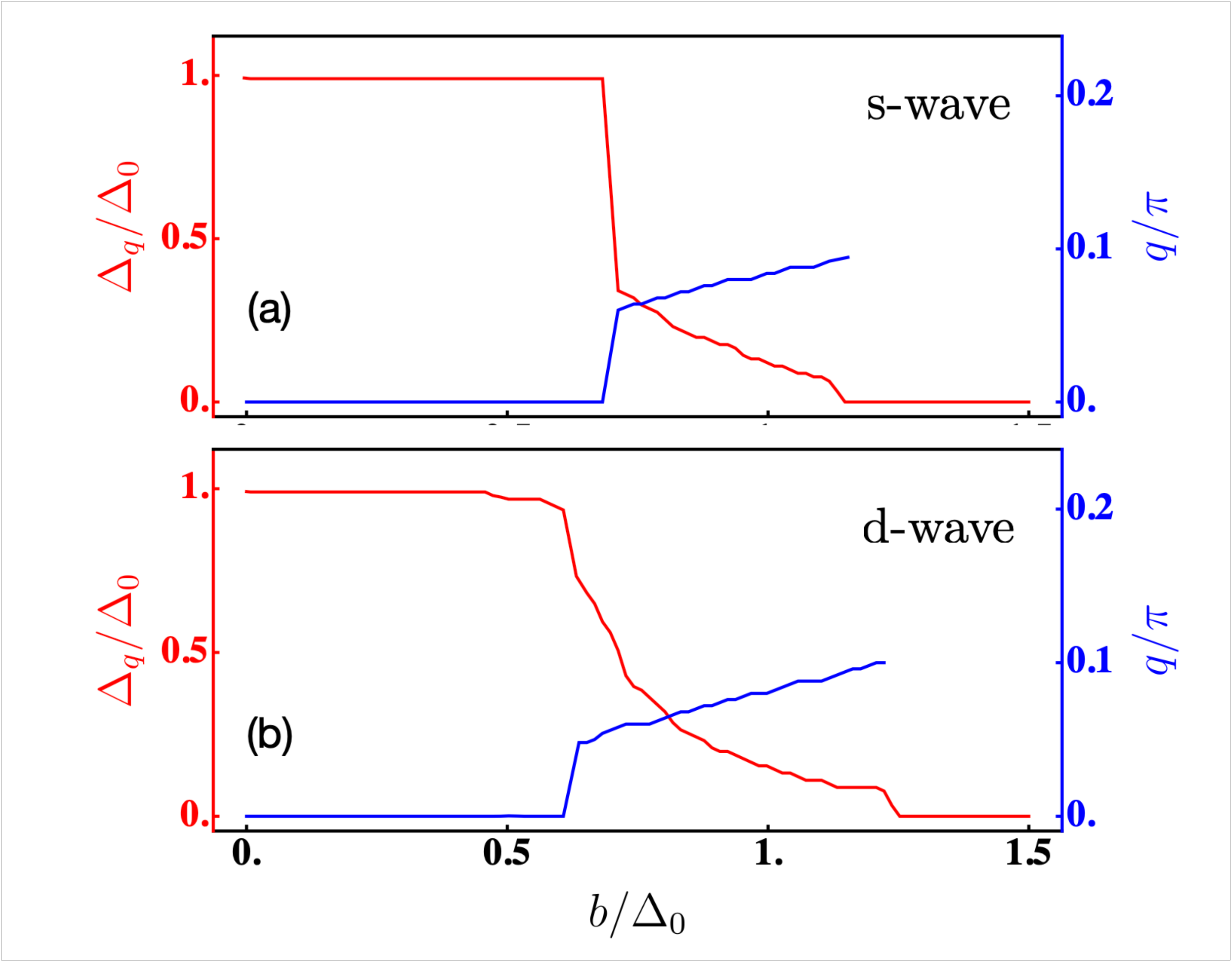}
\caption{Field dependence of $q$-vector (blue lines) and associated gap $\Delta_q(b)$ (red lines)  for Cooper pairing with finite momentum  as function of applied field for s-wave and d-wave gap functions as obtained by minimizing the condensation energy in Eq.~(\ref{eq:gscond}). The critical fields for BCS to FF transition are at $b_c/\Delta_0\simeq 0.7$ for s-wave and $b_c/\Delta_0\simeq 0.6$ for d-wave. Here $\Delta_0=t/2=D_c/8$, $\mu=-2t$ with $D_c$ the half band width. In this and all consecutive figures the energy unit is chosen $t=1$.
}
\label{fig:qvect}
\end{figure}
%

%
\begin{figure}[t]
\includegraphics[width=0.95\linewidth]{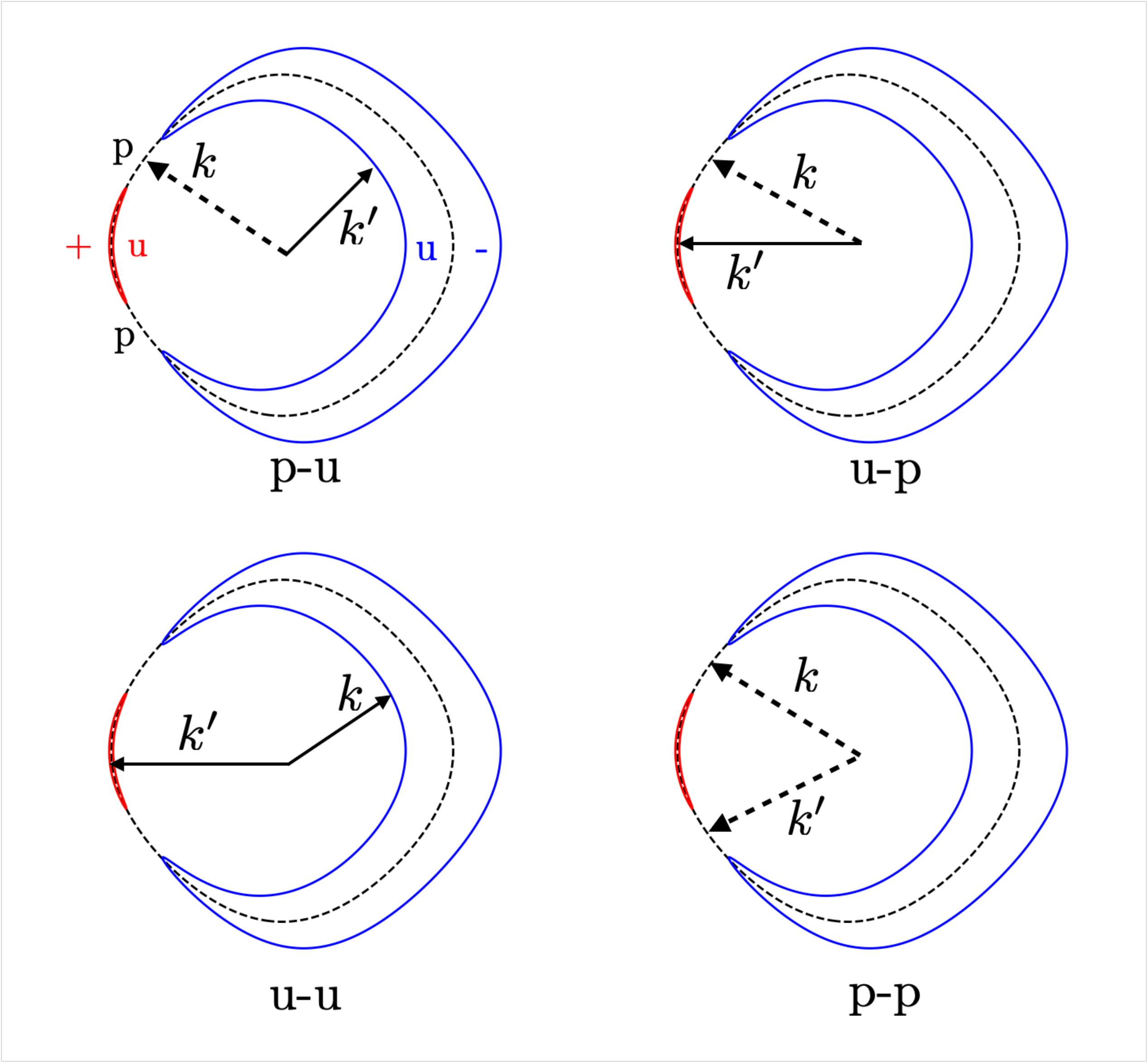}
\caption{Possible excitation processes contributing to the dynamical susceptibility in the FF state at zero temperature (Eq.~(\ref{eq:FFLO-sus0})) for s-wave case. Unpaired (u) Fermi surface sheets are defined by $(\omega=0.1t)$ $-E^-_{\bk\bq}=\omega$ (blue) and $E^+_{\bk\bq}=\omega$ (red) for $b/\Delta_0=0.81, \mu =-2t$. The upper row describes quasiparticle scattering (momenta $\bk$ and $\bk'=\bk+\tbq$) between paired (p) and unpaired (u) Fermi surface segments. The lower row describes quasiparticle destruction or creation either between paired (p-p, dashed arrows)  or unpaired (u-u, full arrows) Fermi surface segments. In the BCS case $(b=0,q=0)$ the red and blue unpaired quasiparticle segments vanish and only the dashed processes of the second row survive.}
\label{fig:process}
\end{figure}
%

In essence  the FFLO superconducting state is characterized by a coherent superposition of paired $(-\bk+\bq\ua, \bk+\bq\da)$ 
and unpaired states whose momenta \bk~belong to different segments of the Fermi surface such that the paired states have a common center of mass (CM) momentum $2\bq$. The formation of this state may be described by a mean field pairing Hamiltonian~\cite{akbari:16}
\be
\bl
H_{\rm SC}
=&
\sum_{\bk}\psi^\dg_{\bk\bq}\hh_{\bk\bq}\psi_{\bk\bq}
+\sum_{\bk}\xi^\da_{\bk+\bq} + N\Bigl(\frac{|\De^0_\bq|^2}{V_0}\Bigr),
\\
\hh_{\bk\bq}
=&
\left(
 \begin{array}{cc}
 \xi_{\bk+\bq\ua}&-\Delta^\bk_{\bq} \\
 -\Delta^{\bk *}_{\bq}& -\xi_{-\bk+\bq\da}
\end{array}\right)
\\
=
&
\;
 (\xi^a_{\bk\bq}+h)\tau_0+
\left(
 \begin{array}{cc}
 \xi^s_{\bk\bq}&-\Delta^\bk_{\bq} \\
 -\Delta^{\bk *}_{\bq}& -\xi^s_{\bk\bq}
\end{array}\right),
\label{eq:Ham}
\el
\ee
where a 2D tight binding (TB) conduction band $\vare_\bk =-2t(\cos k_x +\cos k_y) $ with hopping element $t>0$ and band width $2D_c=8t$ will be used. Furthermore defining $\xi_\bk = \vare_\bk-\mu$ with respect to the chemical potential $\mu$ we will abbreviate the field split conduction band energies $(b=\mu_BB, \sigma =\pm 1~\mbox{or} \ua,\da)$ and its (anti-)symmetrized combinations $s(+)$ and $a(-)$ of  dispersions  shifted by $\pm\bq$ according to
\bea
\xi_\bk^\si=\xi_\bk+\si b; \;\;\; \xi^{s,a}_{\bk\bq}=\fs(\xi_{\bk+\bq}\pm\xi_{\bk-\bq}).
\eea
In Eq.~(\ref{eq:Ham}) $\tau_0$ is the unit in Nambu particle-hole space. Furthermore $\Delta_\bq^\bk$ is the gap function and $\Delta^0_\bq$ its amplitude in the FF state.  The effective interaction strength $V_0$  is defined below in Eq.~(\ref{eq:effint}). The above FF Hamiltonian may be diagonalized by Bogoliubov transformation \cite{cui:06,akbari:16} (which is different for paired and unpaired states) leading to a quasiparticle Hamiltonian
\be
\bl
H_{\rm SC}
=E_G(\bq,\De_\bq)+\frac{1}{2}\sum_{\bk}(|E^+_{\bk\bq}|\al^\dg_\bk\al_\bk+|E^-_{\bk\bq}|\bt^\dg_\bk\bt_\bk).
\label{eq:BCS2}
\el
\ee
The symmetrized form of the Hamiltonian in Eq.~(\ref{eq:Ham}) implies that only the symmetrized dispersions $\xi^s_{\bk\bq}$ will appear in the
Bogoliubov transformation but both  $\xi^s_{\bk\bq}$,  $\xi^a_{\bk\bq}$ and $b$ will be present in the expression for the (positive) quasiparticle excitation energies  $|E^\si_{\bk\bq}|$ $(\sigma=\pm)$ which are given by
\bea
\bl
E^\si_{\bk\bq}
&
=
E_{\bk\bq}+\si(\xi_{\bk\bq}^a+b),
\\
E_{\bk\bq}&=\sqrt{\xi^{s2}_{\bk\bq}+|\De^\bk_\bq|^2}
.
\label{eq:quasienergy}
\el
\eea
Furthermore the ground state energy of the FF phase is obtained as \cite{cui:06,akbari:16}:
\be
\bl
E_G(\bq,\De_\bq)
=
&
N\Bigl(\frac{|\De_\bq|^2}{V_0}\Bigr)-\sum_\bk(E_{\bk\bq}-\xi^s_\bk)
\\
&
+\sum_\bk[E^+_{\bk\bq}\theta(-E^+_{\bk\bq}) +   E^-_{\bk\bq}\theta(-E^-_{\bk\bq})]    
.
\label{eq:gsen}
\el
\ee
\begin{figure*}[t]
\includegraphics[width=0.95\linewidth]{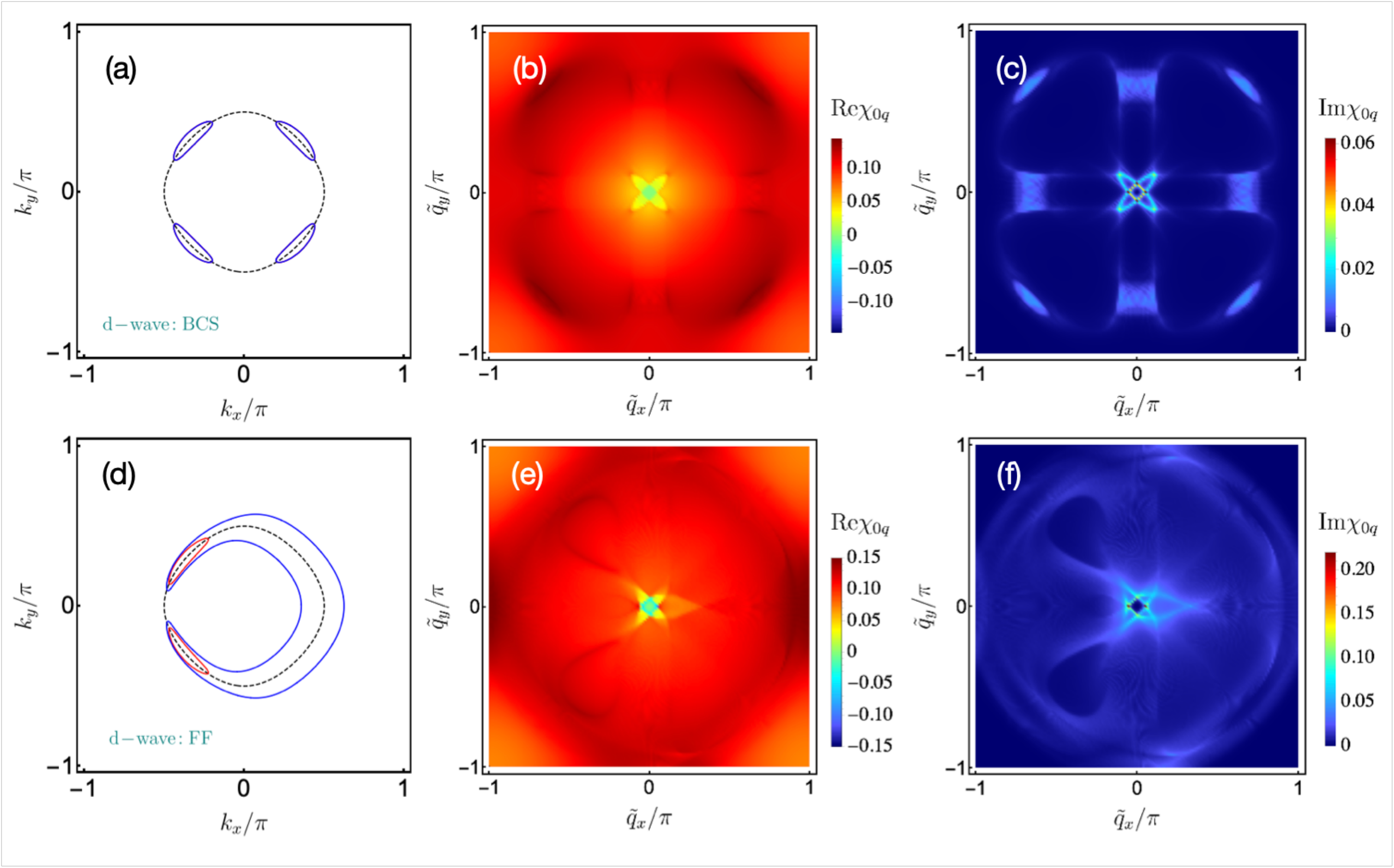}
\caption{Brillouin zone constant- $\omega$ cuts of spectral function, real part  and imaginary part  of dynamical 
susceptibility $\chi_{0\bq}(\tbq,\omega)$ (from left to right, respectively)  for d-wave BCS state (a-c) and d-wave FF state (d-f) and  for $\omega/\Delta_0=0.60$. Parameters $(q,\Delta_q)$ for the FF state are those at $b/\Delta_0=0.63$ in Fig.~\ref{fig:qvect} slightly above the BCS-FF transition. The pockets in (a) are located around nodal positions of the d-wave BCS state. The features for small and large $\tbq$ in (b,c) result from intra-and inter- pocket virtual excitations in (a). In (e,f) the large (blue) unpaired quasiparticle sheets of the FF phase in (d) also contribute to the large $\tbq$ response function. In $\tilde{q}_x$ direction the response function become asymmetric due to the nonzero pair momentum $2q$ oriented along antinodal $\tq_x$ direction (see also Fig.~\ref{fig:suscut}(c,d)).
 }
\label{fig:dcontspec025}
\end{figure*}
There are two possible cases \cite{cui:06}: i) when both $E^\si_{\bk\bq}>0$ one has a stable pair state for momentum  \bk~with CM momentum $2\bq$. When either  $E^+_{\bk\bq}<0$ or  $E^-_{\bk\bq}<0$ the pair state is unstable and one has normal quasiparticle states at \bk~with excitation energy  $|E^\si_{\bk\bq}|>0$.
(The case with both negative  $E^\si_{\bk\bq}$ cannot occur because according to Eq.~(\ref{eq:quasienergy}) their sum must be positive.)\\

We will consider two possible one-dimensional spin-singlet $C_{4v}$- representations $(\Gamma = A_1, B_1)$ for the gap function defined by $\Delta_\bq^\bk=\Delta^0_\bq f_\Gamma(\bk)$ with the form factor $f_\Gamma(\bk)=1$ for the isotropic s-wave and $f_\Gamma(\bk)=\cos k_x-\cos k_y$ in the d-wave cases, respectively.
 The form factors are normalized according to $(1/N)\sum_\bk f^2_\Gamma(\bk)=1$ such that the Brillouin zone (BZ) averaged gaps $\bra\Delta_\bq^{\bk 2}\ket^\fs=\Delta^0_\bq$ are equal in the two cases. Note, however, that the maximum gap modulus at the points $(\pi,0), (0,\pi)$ and equivalents is given by $\Delta^d_\bq=2\Delta^0_\bq$ in the d-wave case which will be used in Sec.~\ref{subsec:spinres}.  The strength $V_0$ of the corresponding pair interactions $V_\Gamma(\bk,\bk')=-V_0f_\Gamma(\bk)f_\Gamma(\bk')$ appearing in Eq.~(\ref{eq:Ham}) is determined via the gap equation for the BCS case $(b=0,q=0)$ as 
\bea
\frac{1}{V_0}=\frac{1}{N}\sum_\bk\frac{f_\Gamma^2(\bk)}{2E_{\bk\bq}},
\label{eq:effint}
\eea
Here the index $\Gamma$ for  $V_0$ has been suppressed.\\

How large the paired and unpaired Fermi segments are depends on the size of CM pair momentum $2q(b)$
and gap size $\Delta_q(b)$ in the FFLO state. They are determined by the minimization of the condensation energy $E_c=E_G-E_G^0$ where $E_G$ is the ground state energy of the superconducting state appearing  in 
Eq.~(\ref{eq:BCS2}) and  $E_G^0=\sum_\bk(\xi_\bk-|\xi_\bk|)$ that of the normal ground state. One obtains \cite{cui:06,akbari:16}:
\be
\bl
E_c(\bq,\De_\bq)
=
&
N\Bigl(\frac{|\De^0_\bq|^2}{V_0}\Bigr)
\!-\!
\sum_\bk(E_{\bk\bq}-|\xi_\bk|)
\!+\!
\sum_\bk(\xi^s_{\bk\bq}-\xi_\bk)
\\&
+\sum_\bk[E^+_{\bk\bq}\theta(-E^+_{\bk\bq}) +   E^-_{\bk\bq}\theta(-E^-_{\bk\bq})]    
.
\label{eq:gscond}
\el
\ee
For each field \bb~the minimum energy state characterized by $(\bq,\Delta_q)$ has to be found numerically from this condensation energy functional. We choose the field $\bb=\mbox{b}\hbz$ and spin quantization axis along z- direction and the FF vector  $\bq=\mbox{q}\hbx$ along x-direction which is the antinodal direction in d-wave case. 

The field dependence of the pair $(\bq,\Delta_q)$ is shown in Fig.\ref{fig:qvect} for $\mu=-2.8t$ where the TB Fermi surface is already distinctly nonspherical (Fig~\ref{fig:process}).
 At a critical field $b^*$ the ground state changes from zero $q$- momentum BCS phase to the FF phase with finite q and reduced gap size $\Delta_q$. The critical field for the nodal d-wave gap is somewhat smaller and the gap reduction less sudden.

%
\begin{figure*}[t]
\includegraphics[width=0.99\linewidth]{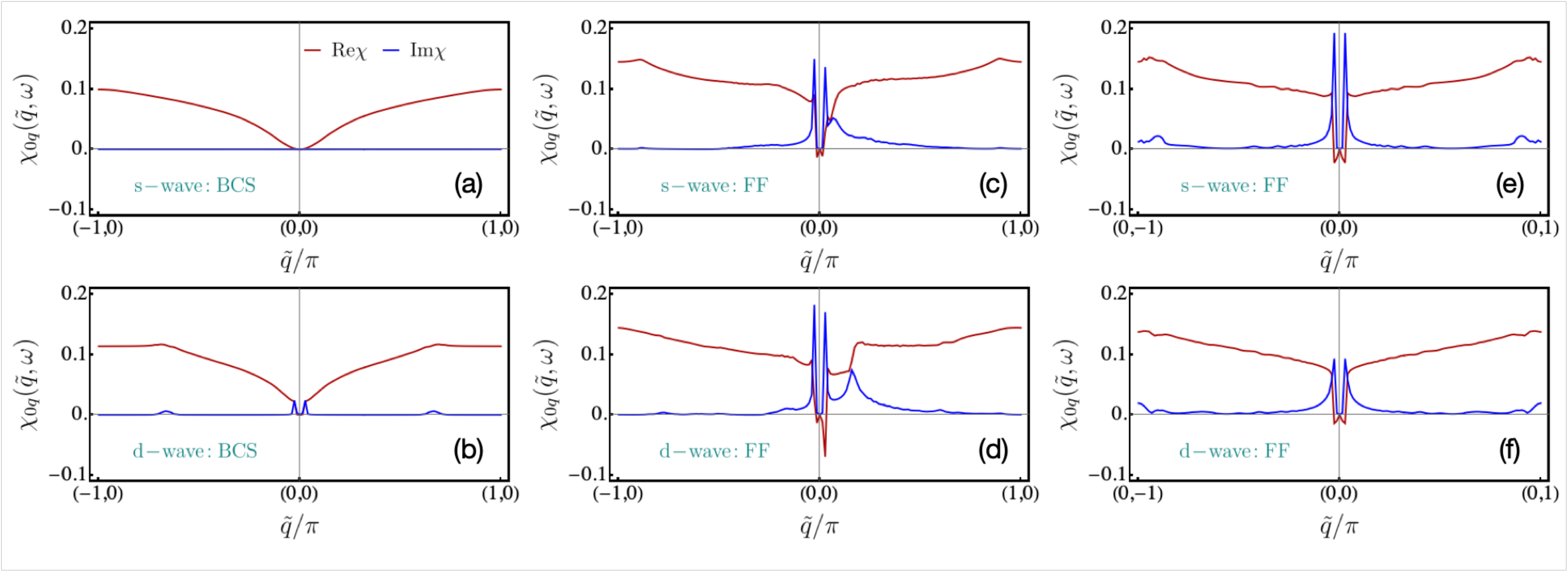}
\caption{Brillouin zone cuts of real and imaginary part of $\chi_{0\bq}(\tbq,\omega)$ $(\omega =0.30\Delta_0)$. (a,b): Along $\tilde{q}_x$ direction for BCS states.
(c,d): Along $\tilde{q}_x$ direction FF state with parameters $(q,\Delta_q)$ corresponding to those at $b/\Delta_0=0.81$  for s-wave case and  $b/\Delta_0=0.63$ for d-wave in Fig.~\ref{fig:qvect}. The asymmetry for small $\tilde{q}_x$  noted in Fig.~\ref{fig:dcontspec025}(e,f) due to the parallel direction of pair momentum $2q$  is clearly visible. (e,f): In the perpendicular $\tilde{q}_y$ direction the reflection symmetry of BCS case (a,b) is preserved. The sharp peaks at small momentum transfer are due to small intra-band tansitions corresonding to small pockets in Fig.~\ref{fig:dcontspec025}(a,d). }
\label{fig:suscut}
\end{figure*}
%

\section{The magnetic response function for the FF state}
\label{sec:FFLO-response}

The static and dynamical spin susceptibility are  important tools to investigate the BCS superconductor using Knight shift and NMR experiments as well as inelastic neutron scattering (INS). It is worthwhile to extend the analysis of this important quantity to the FF phase. An earlier investigation for the d-wave state with fixed pair momentum, coexisting SDW and focused on static results was given in Ref.~\cite{mier:09}. Here we consider the FF state with calculated pair momentum and gap and focus on dynamical properties, in particular with respect to the question of spin resonance behaviour in the FF phase region.

\subsection{Derivation of the general dynamical susceptibility expression}
\label{subsec:dynamical}

\begin{figure}
\includegraphics[width=\linewidth]{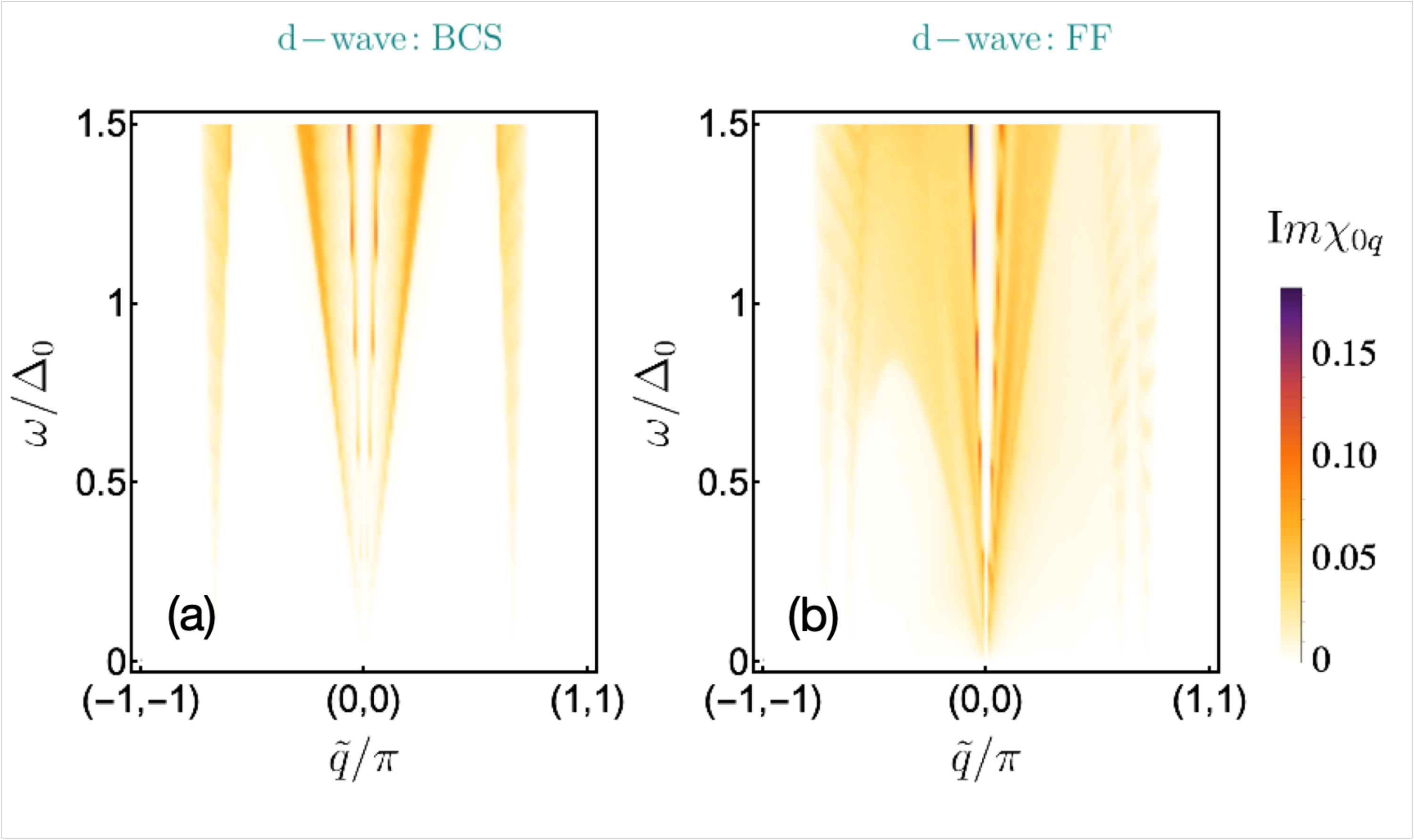}\
\caption{Dispersion of spin excitation spectrum along $(1,1)$ direction for BCS (a) and FF (b) d-wave case with $b/\Delta_0=0.63$. (cf. Fig.~\ref{fig:dcontspec025}(c,f)).  The central branch is due to intrapocket and the outer branch (the two parts are connected at larger $\omega$) due to inter-pocket excitations. In the FF case (see Fig.~\ref{fig:dcontspec025}(d)) one small pocket pair is lost and therefore the outer branch is  suppressed (b) whereas the intensity of the inner branch becomes asymmetric in accordance with Fig.~\ref{fig:suscut}(c,d). 
}
\label{fig:specdisp}
\end{figure}

The bare magnetic response function of a superconductor obtained from the bubble diagram without vertex corrrections is given by \cite{michal:11}
\be
\bl
\chi^{\al\al}_{0\bq}(\tbq,\num) 
&=
\\
&
\hspace{-0.85cm}
-\frac{T}{4}\frac{1}{N}\sum_{\bk n\si\si'}\si^\al_{\si\si'}\si^\al_{\si'\si}
{\rm Tr}_\tau[\hG_\bq(\bk,\om)\hG_\bq(\bk',\om')].
\el
\ee
Here $\tau$ is the index in particle-hole space of the Nambu Green's functions $\hG_\bq(\tbq,\om)$ of the FF state. Furthermore $\tbq=\bk'-\bk$ is the momentum transfer and $\bq$ the (half-) pair momentum. We perform the spin sum which is isotropic (independent of $\alpha=x,y,z)$ for singlet pairs and use the explicit form of the Green's function matrix
\be
\bl
\hG_{\bq}(\bk,\om)
&=
(\om-\hh_{\bk\bq})^{-1}
\\
&=
\frac{1}{D_{\bk\bq}(\om)}
\left(
 \begin{array}{cc}
 \om+\xi^\da_{\bk-\bq}   &-\Delta^\bk_{\bq} \\
 -\Delta^{\bk *}_{\bq}& \om-\xi^\ua_{\bk+\bq}
\end{array}
\right).
\label{eq:greenmat}
\el
\ee
with 
\bea
\bl
D_{\bk\bq}(\om)
&=(\om-\xi^\ua_{\bk+\bq})(\om+\xi^\da_{\bk-\bq})-|\De_\bq|^2
\quad
\\
&=(\om-E^+_{\bk\bq})(\om+E^-_{\bk\bq}).
\el
\eea
Then we obtain in closed form, suppressing spin index $\alpha$ from now on: 
\be
\bl
&\chi_{0\bq}(\tbq,\num) 
=
-\frac{T}{4}\frac{1}{N}\sum_{\bk,n}
\\ 
&
\quad
\frac
{
(\om-\zeta_{\bk\bq})(\omp-\zeta_{\bk'\bq})+\xi_{\bk\bq}^s\xi_{\bk'\bq}^s+\Delta_\bq^\bk\Delta_\bq^{\bk'}
}
{(\om-E^+_{\bk\bq})(\om+E^-_{\bk\bq})(\omp-E^+_{\bk'\bq})(\omp+E^+_{\bk'\bq})
}.
\el
\ee
Carrying out the summation over the Matsubara frequencies $\omega_n$ and analytically continuing to the real axis according to $\num\rightarrow \omega+i\eta$ a lengthy calculation leads to the
final result:
\be
\bl
&
\chi_{0\bq}(\tbq,\omega) =\chi^{sc}_{0\bq}(\tbq,\omega)+\chi^{ac}_{0\bq}(\tbq,\omega)
=
\\
&
\frac{1}{2N}
\sum_\bk 
\Bigl\{
\tC^\bq_+(\bk\bk') \times\\&
\Bigl[
\frac{f(E^+_{\bk'\bq})-f(E^+_{\bk\bq})}{\omega-(E^+_{\bk'\bq}-E^+_{\bk\bq})+i\eta}
-\frac{f(E^-_{\bk'\bq})-f(E^-_{\bk\bq})}{\omega+(E^-_{\bk'\bq}-E^-_{\bk\bq})+i\eta}
\Bigr]+
\\
&
\tC^\bq_-(\bk\bk')
\times
\\&
\Bigl[
\frac{1-f(E^-_{\bk'\bq})-f(E^+_{\bk\bq})}{\omega+(E^-_{\bk'\bq}+E^+_{\bk\bq})+i\eta}+
\frac{f(E^+_{\bk'\bq})+f(E^-_{\bk\bq})-1}{\omega-(E^+_{\bk'\bq}+E^-_{\bk\bq})+i\eta}
\Bigr]
\Bigr\},
\label{eq:FFLO-sus}
\el
\ee
where $f(E)=(\exp(E/T)+1)^{-1}$ is the Fermi function. The last two terms may also be written differently  by using $1-f(E)=f(-E)$.
Here the generalized superconducting coherence factors of magnetic response for the FF phase are given by
\bea
\tC^\bq_\pm(\bk\bk')=\fs\Bigl[
1\pm \frac{\xi_{\bk\bq}^s\xi_{\bk'\bq}^s+\Delta_\bq^\bk\Delta_\bq^{\bk'}}
{E_{\bk\bq}E_{\bk'\bq}}\Bigr].
\label{eq:FFLO-coh}
\eea
Note that only the q-symmetrized dispersions $\xi^s_{\bk\bq}$ (directly and implicitly in $E_{\bk\bq}$) appear in the coherence factors.
The above magnetic response function for the FF phase reduces to the well known result \cite{bulut:96,norman:00,michal:11}
in the BCS limit $b,q=0$ which is given in Appendix \ref{sec:BCS-response} for comparison. We note that the sequence in which quasiparticle dispersions $E^\sigma_{\bk\bq}$ appear in Eq.~(\ref{eq:FFLO-sus}) could not be guessed heuristically from the $b,q=0$ BCS expression in Eq.~(\ref{eq:BCS-sus}).  If we restrict to the case
where $\bk$, $\bk'$ lie both in the segment with paired states (i.e. $E^\pm_{\bk\bq}>0, E^\pm_{\bk'\bq}>0$)  then the terms in Eq.~(\ref{eq:FFLO-sus})
may be consecutively interpreted as: quasiparticle scattering ($\chi^{sc}$) (first two terms) and sum ($\chi^{ac}$) of pair annihilation (third) and pair creation (fourth) terms.
For general $\bk$, $\bk'$~one has to consider processes involving quasiparticles from the paired (p) as well as the unpaired (u) Fermi surface
segments. To simplify matters in this general case we consider the zero temperature limit when the Fermi function may be expressed by the step 
function according to  $f(E)=1-\Theta(E)=\Theta(-E)$. Then we obtain
\be
\bl
&
\chi_{0\bq}(\tbq,\omega) =
\frac{1}{2N}
\sum_\bk 
\Bigl\{
\\&\;
\tC^\bq_+(\bk\bk')\times
\\&\;
\Bigl[
\frac{\Theta(E^+_{\bk\bq})-\Theta(E^+_{\bk'\bq})}
{\omega-(E^+_{\bk'\bq}
\!-\!
E^+_{\bk\bq})+i\eta}
\!-\!
\frac{\Theta(E^-_{\bk\bq})
\!-\!
\Theta(E^-_{\bk'\bq})}
{\omega
\!+\!
(E^-_{\bk'\bq}
\!-\!
E^-_{\bk\bq})+i\eta}
\Bigr]+
\\
&\;
\tC^\bq_-(\bk\bk')\times
\\&\;
\Bigl[
\frac{\Theta(E^+_{\bk\bq})-\Theta(-E^-_{\bk'\bq})}{\omega+(E^-_{\bk'\bq}+E^+_{\bk\bq})+i\eta}+
\frac{\Theta(-E^+_{\bk'\bq})-\Theta(E^-_{\bk\bq})}{\omega-(E^+_{\bk'\bq}+E^-_{\bk\bq})+i\eta}
\Bigr]
\Bigr\}
.
\label{eq:FFLO-sus0}
\el
\ee
If we look at the numerators of the four terms in this equations we realize that the first two correspond
to quasiparticle scattering processes $\bk\leftrightarrow\bk'$ from paired to unpaired  FS segments and 
vice versa (p-u,u-p). whereas the third and fourth term are quasiparticle annihilation and creation respectively,
containing only processes between the paired (p-p) or unpaired (u-u) segments. The various possible contributions are illustrated pictorially in the spectral plot of Fig.~\ref{fig:process} and the resulting bare response function is shown in Figs.~\ref{fig:dcontspec025},\ref{fig:suscut},\ref{fig:specdisp} and  discussed in Sec.~\ref{sec:discussion}.

\subsection{The static susceptibility, Knight shift and Yosida function in the FF state}
\label{subsec:static}

Although the  general wave vector $\tbq$- dependent static  susceptibility  cannot be directly measured, it is interesting to derive its formal structure. Setting $\omega =0$ in Eq.~(\ref{eq:FFLO-sus}) we obtain for $\chi_{0\bq}(\tbq) \equiv \chi_{0\bq}(\tbq,0) $:
\be
\bl
\chi_{0\bq}(\tbq) 
=\!
\frac{1}{2N}\sum_{\bk\si} 
&
\Bigl\{
\tC^\bq_+(\bk\bk')
\frac{\tanh\frac{\beta}{2}E^\si_{\bk'\bq}-\tanh\frac{\beta}{2}E^\si_{\bk\bq}}
{E^\si_{\bk'\bq}-E^\si_{\bk\bq}}
\\
&
+
\tC^\bq_-(\bk\bk')
\frac{\tanh\frac{\beta}{2}E^\si_{\bk'\bq}+\tanh\frac{\beta}{2}E^{\bar{\si}}_{\bk\bq}}
{E^\si_{\bk'\bq}+E^{\bar{\si}}_{\bk\bq}}
\Bigr\}.
\label{eq:FFLO-susstat}
\el
\ee
where $\bar{\sigma}=-\sigma$.
This may be further simplified for the homogeneous susceptibility with $\tbq=0$. Using $\tC_+^\bq(\bk\bk)=1$ and $\tC_-^\bq(\bk\bk)=0$  it can be derived as 
\be
\chi_{0\bq}(0)=\frac{1}{2N}\sum_{\bk\si}
\Bigl(-\frac{\partial f}{\partial E^\si_{\bk\bq}}\Bigr)
=\frac{\beta}{4}\frac{1}{2N}\sum_{\bk\si}\Bigl(\frac{1}{\cosh^2\frac{\beta}{2}E^\si_{\bk\bq}} \Bigr)
.
\ee
This $\tbq=0$ static susceptibility is therefore proportional to the T-averaged DOS of quasiparticles at a given temperature and this determines the T-dependence of the  Knight shift in an NMR experiment in the superconductor. Usually, in the zero-field BCS singlet superconducting state this quantity contains information on the nodal structure of the SC gap function. In the FF state, however it is also determined by the normal quasiparticles in the depaired momentum space segments and is  influenced by them. This problem has also been considered with a different quasiclassical method for spatially inhomogeneous d-wave state \cite{vorontsov:06}.\\
%
%
%
\begin{figure}
\includegraphics[width=0.990\linewidth]{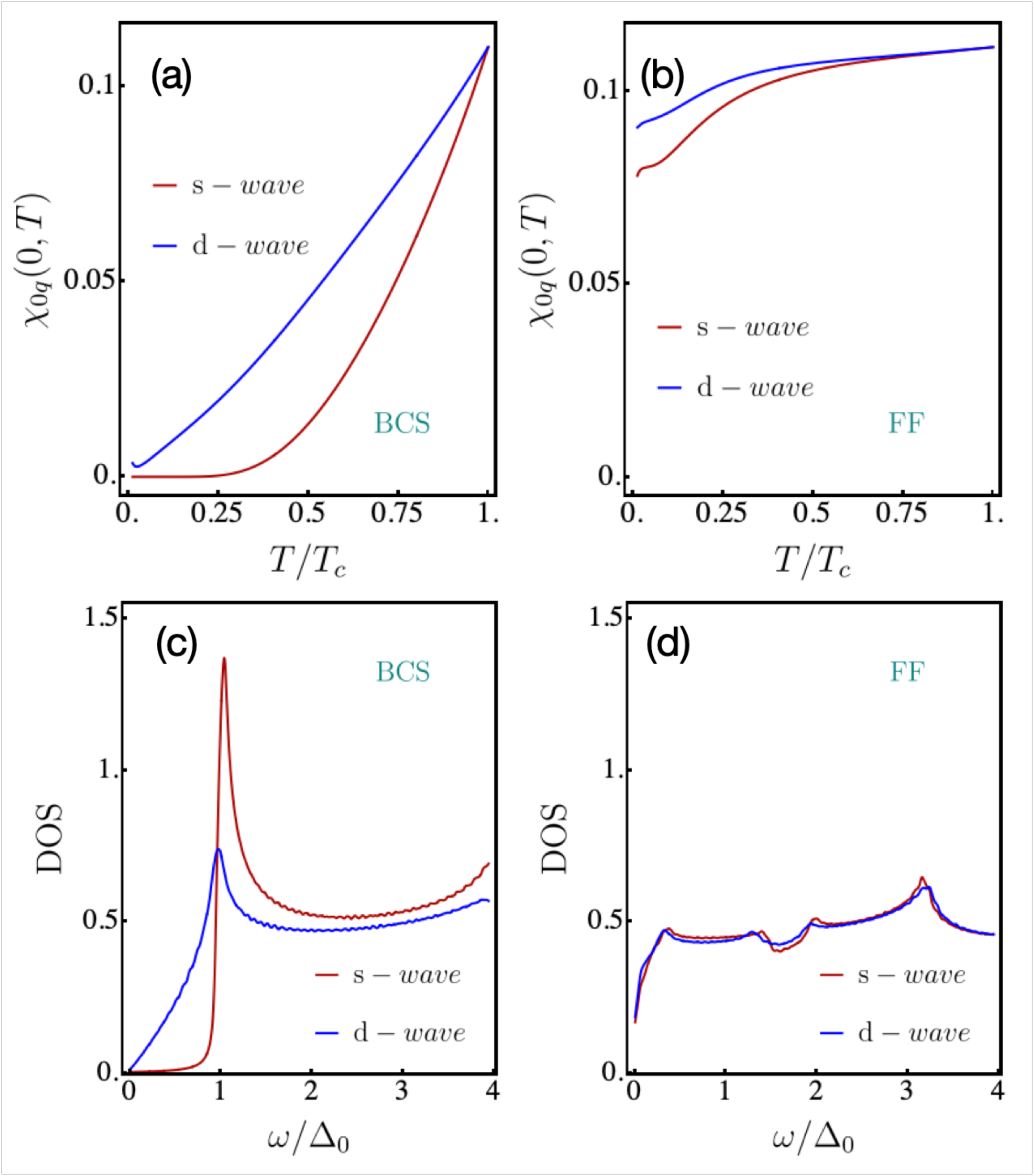}\
\caption{Temperature dependence of static susceptibility (units $1/t$) or Yosida function for (a) BCS cases and (b) FF phase ($b/\Delta_0=0.81, T_c(b)/\Delta_0=0.5)$ as function of reduced temperature $t_r=T/T_c(b)$. In (a) the exponential and power law decay for s- and d-wave case are clearly distinct. In the FF phase of (b) unpaired quasiparticles appear in both cases leading to large values even at  low temperatures. The corresponding quasiparticle DOS is shown in (c,d).}
\label{fig:yosida}
\end{figure}
%
In the parabolic band approximation (for $\mu\ll D_c$) with a 2D DOS $N_0=m^*k_F/2\pi$ and effective mass $m^*=2/D_c$ and Fermi vector $k_F=(2m^*\mu)^\fs$ this may be written as
\be
\bl
\chi_{0\bq}(0,T)
&=
N_0\sum_\si Y^\si_q(T);\;\;\; 
\\
Y^\si_q(T)
&=
\int\frac{d\theta_\bk}{2\pi}\Bigl[
\frac{1}{4\pi}\int\frac{d\xi}{\cosh^2\frac{\beta}{2}E^\si_{\bk\bq}}\Bigr]
,
\label{eq:yosida}
\el
\ee
where $Y^\si_q(T)$ is the generalized Yosida function \cite{mineev:99} that describes the temperature dependence of the NMR Knight shift of the singlet superconductor in the FF phase. For plotting the temperature dependence of the homogeneous static susceptibillity we use a phenomenological temperature dependence of the FF gap $\Delta_q$ given by the expression $\Delta_q(t_r)=\Delta_q\tanh[1.74\sqrt{\frac{1-t_r}{t_r}}]$ where $t_r=T/T_c(b)$ is the reduced temperature referenced to the relevant $T_c(b)$. The comparison of $\chi_{0\bq}(0,T)$ in the BCS $(q=0)$ and FF $(q\neq 0)$ case in the interval $t_r\in [0,1]$ is shown in Fig.~\ref{fig:yosida},
together with corresponding DOS curves, and discussed in Sec.~\ref{sec:discussion}.

\subsection{Spin resonance excitation in the d-wave FF state}
\label{subsec:spinres}

It is known from many examples, in particular from the f-based heavy fermion superconductors \cite{thalmeier:16, eremin:08, *akbari:21, *akbari:12, stock:08,raymond:12} but also from high-T$_c$ \cite{eschrig:06} and Fe- pnictide \cite{korshunov:08,inosov:10}  compounds that the dynamic magnetic response for unconventional gap symmetry can exhibit the spin resonance excitations within the superconducting gap of the BCS phase, i.e., at a resonance frequency $\omega_r/2\Delta_d<1$ where $\Delta_d=2\Delta_0$ (Sec.~\ref{sec: FFLO}) is the maximum d-wave BCS gap value. For a half filled $(\mu=0)$ TB band in the d-wave case it is centered around the zone boundary vector $\tbq\equiv\bQ=(\pi,\pi)$. The resonance formation is due to the peculiar step-like behaviour of  ${\rm Im}\chi_0(\tbq,\omega)$  and associated peak in  ${\rm Re}\chi_0(\tbq,\omega)$ at the threshold energy of quasiparticle excitations (more precisely the threshold is $\omega_r(\tbq)< \min_{\bk\in\mbox{FS}}(|\Delta_\bk |+|\Delta_{\bk+\tbq}|)$ rather than the upper limit $2\Delta_d$). The resonance is observed only  for unconventional gap functions which change sign under translation by the wave vector $\tbq$. (see Appendix \ref{sec:BCS-response}). 

\begin{figure}
\includegraphics[width=0.990\linewidth]{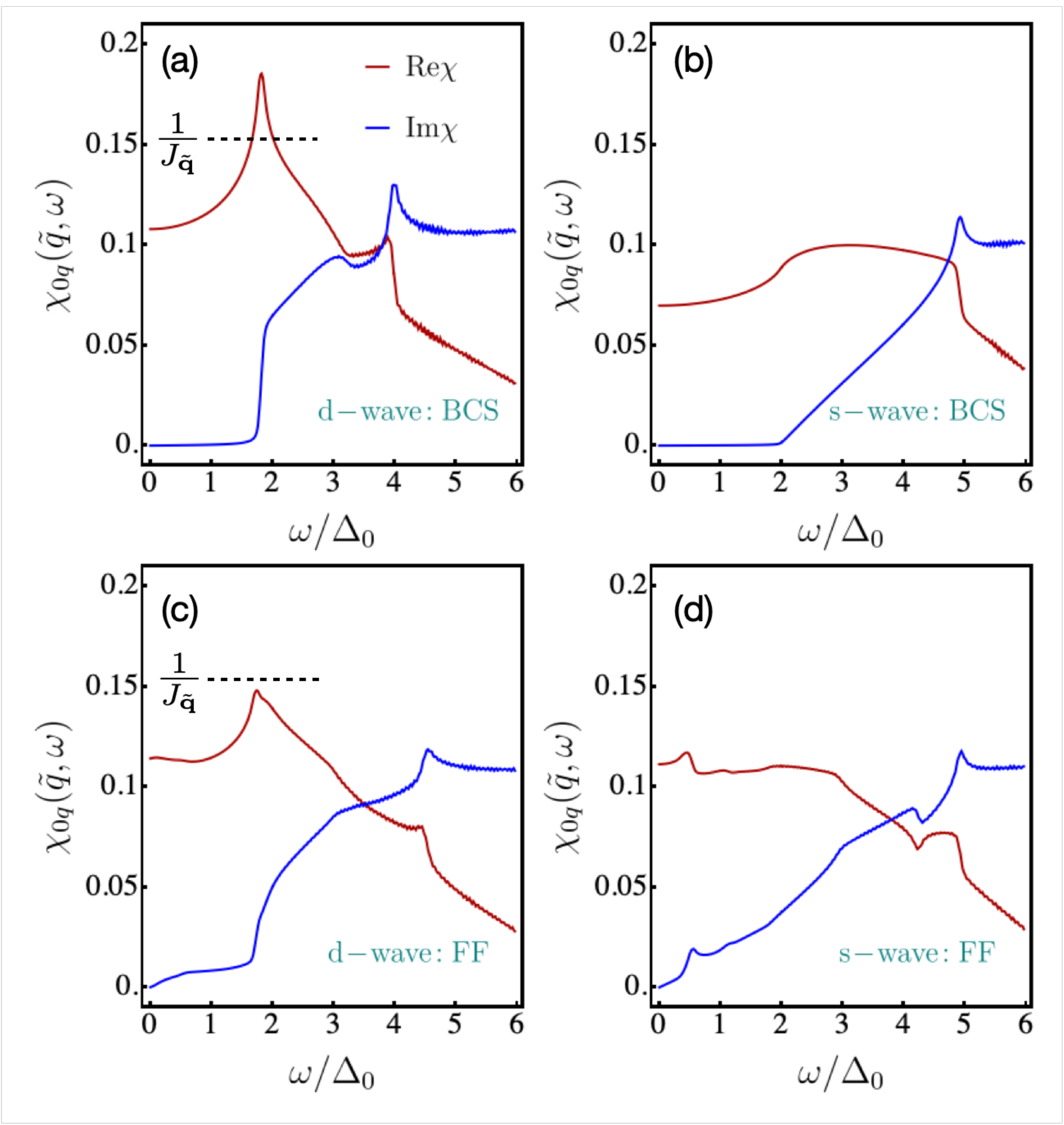}
\caption{Frequency dependence of susceptibility at momentum transfer $\tbq=(0.4\pi,0.4\pi)$. (b,d): s-wave BCS and FF-cases. (a,b): d-wave BCS and FF- phases. In FF phase the $(q,\Delta_q)$ parameter correspond to $ b=0.71\Delta_0$ for s-wave and $ b=0.63\Delta_0$ for d-wave in Fig.~\ref{fig:qvect}. (a,b): In the s-wave case the frequency dependence is featureless but in d-wave BCS case (a)  a step-like increase in imaginary part and peak in real part appear due to behaviour of d-wave coherence factor $\tC_-(\bk,\bk+\tbq)$. The peak in the real part can lead to the spin resonance according to Eq.~(\ref{eq:resonance}) for suitable $1/J_{\tbq}\simeq 0.16$. In the FF phase the peak is strongly suppressed and so will be the resonance.}
\label{fig:susom}
\end{figure}

Here we investigate how the spin resonance appearance is modified in the FF phase of a d-wave superconductor.  Firstly if the external field is appreciably larger than $b^*$ the reduced gap $\Delta^d_q(b)$  in the FF phase (Fig.~\ref{fig:qvect}) will push any perspective surviving resonance to an energy $\omega_r/2\Delta^d_q < 1$ in this case. But the coherence factors and the segmentation of Fermi surface sheets should also influence the resonance features of the collective response.
For this purpose we consider the  RPA susceptibility $\chi^{\rm RPA}_\bq(\tbq,\omega)=[1-J_{\tbq}\chi_{0\bq}(\tbq,\omega)]^{-1}\chi_{0\bq}(\tbq,\omega)$, assuming that low energy quasiparticles have an effective spin exchange interaction given by $J_{\tbq}$. We mention again that $\tbq$ is the momentum transfer in the magnetic response function wheras \bq~is the overall (half-) momentum of  Cooper pairs in the FF phase. Then the dynamical structure function $S(\tbq,\omega)$ investigated in INS is proportional to the imaginary part of the collective RPA susceptibility as given explicitly by
\be
\bl
{\rm Im}\chi^{\rm RPA}_\bq(\tbq,\omega)
&=
\\
&
\hspace{-0.75cm}
\frac{{\rm Im}\chi_{0\bq}(\tbq,\omega)}
{(1-J_{\tbq}
{\rm Re}\chi_{0\bq}(\tbq,\omega))^2+J_{\tbq}^2
(
{\rm Im}
\chi_{0\bq}({\tbq},\omega))^2}.
\label{eq:imchi}
\el
\ee
The INS cross section will therefore develop a peak at $\omega_r$ when the resonance condition 
\bea
\frac{1}{J_{\tbq}}={\rm Re}\chi_{0\bq}(\tbq,\omega_r)
\label{eq:resonance}
\eea
is fulfilled for sufficiently small imaginary part at this frequency. From the d-wave BCS case (see Appendix \ref{sec:BCS-response}) it is known \cite{thalmeier:16} that for perfect nesting FS $(\mu=0)$ when  $\Delta_{\bk+\bQ}=-\Delta_\bk$ for all $\bk$ the resonance peak appears at \bQ. In the FF phase the the resonance will appear at a frequency $\omega_r(\tbq,b) < 2\Delta^d_q(b)$ which depends on the field implicitly through the FF momentum $q(b)$ and gap amplitude $\Delta^d_q(b)=2\Delta_q(b)$ (Fig.~\ref{fig:qvect}(b)). Examples of the frequency dependence of the bare susceptibility $\chi_{0\bq}(\tbq,\omega)$ and the issue of the resonance condition are presented in Figs.~\ref{fig:susom},\ref{fig:spinres} and discussed below.

%
%

\section{Discussion of numerical results: the dynamical spectral functions and static magnetic response}
\label{sec:discussion}

\begin{figure}
\includegraphics[width=1.0\linewidth]{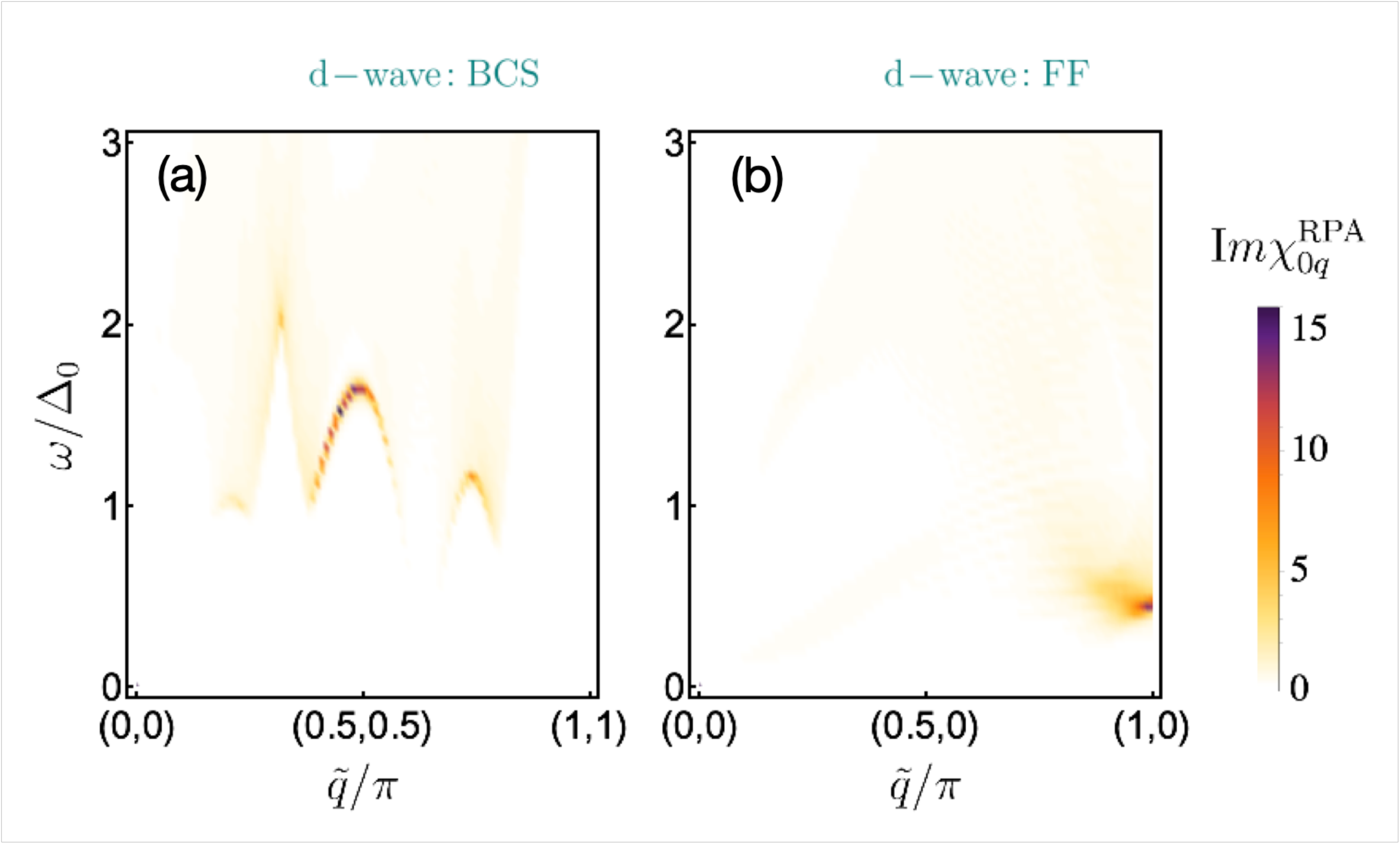}
\caption{Change of spin resonance peak characteristics when moving from BCS (a) to FF (b) phase in d-wave case.
In BCS case (a) the resonance appears prominently in the momentum space region around  $\tbq=(0.5\pi,0.5\pi)$ showing some
dispersion. In the FF case the resonance moves to the zone boundary $\tbq=(\pi,0)$ with lower intensity  and strongly localized.}
\label{fig:spinres}
\end{figure}

One way to observe the profound influence of finite momentum pairs on the magnetic response function is comparison of spectral function (constant frequency cuts of the dispersions $|E^\si_{\bk\bq}|$ and the real and imaginary parts of the dynamical susceptibility. This is shown in Fig.~\ref{fig:dcontspec025} for the d-wave BCS (a-c) and FF phases (d-f). In the former the nodal pockets around $(\pi,\pi)$ direction in (a)  lead to corresponding susceptibility features which result from intra-pocket (small momentum transfer $\tbq$) and inter- pocket (large momentum transfer $\tbq$) excitations. The constant-frequency cuts of the susceptibility in the BZ exhibit the fourfold symmetry of the spectral function. In contrast, in the FF phase (d-f) the finite $q$ momentum parallel to the antinodal $\tilde{q}_x$ direction destroys the reflection symmetry with respect to this axis. This is seen in the spectral function (d) by the merging of $\tilde{q}_x>0$  pockets into a large  combined sheet (blue) whereas the $\tilde{q}_x<0$  pockets survive. It is particularly evident for the small momentum transfer peaks in the imaginary part (f).
 For larger momentum transfer the features localized at $(\pi,0),(\pi,\pi)$ and equivalents merge into a weak ring-shaped structure that also lacks reflection symmetry along $\tilde{q}_x$. Note that in the real part the symmetry breaking is primarily evident from the shifting of the maximum from $(1,1)$ direction in (b) to $(1,0)$ direction parallel to q in (e). This has implications for the spin resonance formation discussed below in connection with Fig.~\ref{fig:spinres}.\\

The $C_{4v}$  symmetry breaking of the magnetic spectrum becomes even more evident by comparing $\chi_{0q}(\tbq,\omega)$ for $\tbq$- directions parallel and perpendicular to the pair momentum direction $(1,0)$ as shown in Fig.~\ref{fig:suscut} for a constant $\omega=0.3\Delta_0$. It presents cuts along $(1,0)$ and $(0,1)$ directions for BCS (a,b) and FF (c,d);(e,f) for s-, d- wave gaps, respectively. In the BCS case the cuts along the two momentum axes are equivalent and both symmetric with respect to reflection $(\tq_x\rightarrow -\tq_x)$. It still holds for the perpendicular direction $(01)$ (e,f). For the BCS case due to $\omega\ll\Delta_0$ the s-wave spectrum (imaginary part, blue) vanishes whereas only small peaks appear for the d-wave case due to intra- and inter- nodal low energy excitations. In the FF-phase more low energy quasiparticle excitations are possible due to depaired regions and pronounced spectral peaks for both s- and d- wave case appear. Significantly in the FF phase the spectrum and the associated real part (red) become quite asymmetric for $(1,0)$ direction due to the finite parallel pair momentum 2q. One may conjecture that this symmetry breaking of the magnetic excitation spectrum in the FF phase may be observable in constant- $\omega$ cuts obtained from INS.
One should note, however, that from such observation it does not seem possible to obtain a direct measure of pair momentum $2q$ which enters into the magnetic spectrum in a rather complicated manner via the quasiparticle excitation energies.\\ 

A complementary way to view the change of the magnetic excitation spectrum across the critical field $b^*$ separating BCS and FF phase is presented in Fig.~\ref{fig:specdisp} for the d-wave case. Here the dispersion of the excitation continuum in the $(\tbq,\omega)$ plane is shown for the $\tbq\parallel (1,1)$ direction. In the BCS phase the two $\Gamma (0,0)$ centered branches corresponding to the butterfly in Fig.~\ref{fig:dcontspec025}(c) and one branch corresponding to the lens close to M$(\pi,\pi)$ appear symmetrically. At higher energy they are merging. In the FF-phase the innermost branch is still present but the intensity is asymmetric under $\tq_x\rightarrow -\tq_x$ whereas the outer branches become blurred into a low intensity continuum.\\

Besides the dynamics  discussed above which is relevant for INS the static homogeneous susceptibility is also an important quantity because it is proportional to the Knight shift observed in NMR experiments \cite{mineev:99}. We show its temperature dependence in Fig.~\ref{fig:yosida}(a,b) using the model parameters described in Sec.~\ref{subsec:static}. In the BCS case one can see the well known distinction between exponential s-wave decay and d-wave power law behaviour of the Yosida function. In the FF phase a finite quasiparticle DOS appears in both cases (Fig.~\ref{fig:yosida}(c,d)) due to the depaired momentum space regions leading to a finite low temperature susceptibility and Knight shift. The distinction between s- and d-wave case is then much less pronounced.\\

Finally we discuss the possibility of the spin-resonance phenomenon in the d-wave case as outlined in Sec.~\ref{subsec:spinres}. We present the real and imaginary parts of the susceptibility for a constant $\tbq$- vector as function of frequency in Fig.~\ref{fig:susom}. In the s-wave case (b,d) due to the lack of sign change property of the gap function the real and imaginary part in BCS as well as FF phase are featureless and the
spin resonance cannot form. For the d-wave case we choose $\tbq=(0.4\pi,0.4\pi)$ in the vicinity of the maximum in Fig.~\ref{fig:dcontspec025}  which connects states of the $(b=0)$ nodal Fermi surfaces which lie on opposite sides of the nodal $(1,1),(-1,1)$ lines leading to a sign change in the gap function (Sec.~\ref{subsec:spinres}). Therefore a step in the imaginary part and associated peak in the real part at the threshold  energy (Fig.~\ref{fig:susom}(a)) appear. If the spin exchange interaction between quasiparticles is sufficiently large, i.e., if $1/J_{\tbq}$ is sufficiently small as indicated in this figure the resonance condition of Eq.~(\ref{eq:resonance}) will be satisfied and a pronounced resonance peak in the collective response spectrum (Eq.~(\ref{eq:imchi})) of the d-wave BCS state is created in the vicinity of this wave vector. When we enter the FF phase the Fig.~\ref{fig:susom}(d) shows that the peak in the real part at the threshold is much diminished such that the resonance condition may no longer be fulfilled. Therefore the collective response in the FF phase will be subdued and/or moved to a different wave vector $\tbq$. For a suitably sized $1/J_{\tbq} =0.16$ this may happen as illustrated in Fig.~\ref{fig:spinres}.
 In the BCS case $(b=0,q=0)$ the dispersive resonance appears around $(0.5\pi,0.5\pi)$ (Fig.~\ref{fig:spinres}(a)) close to the maximum of the real part in Fig.~\ref{fig:susom}(a). In the FF case this resonance peak is suppressed and it reappears at the zone boundary $\tbq=(\pi,0)$ at a rather lower energy and a more localized intensity (Fig.~\ref{fig:spinres}(b)). In any case the quasiparticle sheets in FF phase which exhibit less distinct sign change properties like in BCS case for the gap function and consequently lead to a less favorable situation for the spin resonance formation.
%
%
\section{Conclusion and Outlook}
\label{sec:conclusion}

We have investigated the dynamical magnetic response in the Fulde-Ferrell superconducting phase characterized by Cooper pairs with finite center of mass momentum 2q.  We have derived general analytical expressions for the dynamical magnetic susceptibility and shown that it reduces to the well known result for the BCS phase with $q=0$.
In the latter only excitations between gapped quasiparticles on the completely paired Fermi surface contribute. In the FF phase excitations between paired and unpaired quasiparticle states contribute as well where the latter are gapless.

As an explicit model we consider a single orbital tight binding band and s- and d- wave singlet pairs. By minimization of the total condensation energy we determine the field dependence of (half-) pair momentum $q(b)$ and associated FF gap $\Delta_q(b)$ as input quantities  for the two quasiparticle branches $E^\si_{\bk\bq}$ that determine the magnetic response function. 

We find that  the presence of finite momentum pairs breaks the fourfold $C_{4v}$ symmetry of the susceptibility in the square BZ with only twofold rotations and reflections perpedicular to the (half-) pair momentum q remaining. This is particularly evident for the d-wave case in the small momentum transfer $\tbq$- regime and 
should be observable by constant -$\omega$ scans as well as in the dispersive continuum excitations in the $(\tbq,\omega)$ plane accessible by INS in the FF phase. 

The static susceptibility determines the Knight shift and its temperature dependence shows the well known distinction between exponential and power-law dependence for s- and d- wave cases, respectively, at low temperature for the BCS phase. It was found that the gapless unpaired states appearing in the FF phase lead to a rapid appearance of a large residual low temperature Knight shift. 

We also considered the fate of a possible in-gap collective spin resonance that may appear in a d-wave BCS state when entering the FF phase. We observe that the condition for the resonance formation, i.e. the presence of a large peak in the real part of the dynamical susceptibility is harder to fulfil in the FF case. Therefore one may expect a suppression of the resonance and/or a shift to different wave vectors in this state. This would be an interesting subject to explore with inelastic neutron scattering. 



\appendix

\section{Limiting BCS case of the response function}
\label{sec:BCS-response}

It is worthwhile to see whether the generalized FF dynamical response derived in Eq.~(\ref{eq:FFLO-sus}) reduces to the well known result for the BCS case. First we reformulate $\chi^{sc}_{0q}$ in Eq.~(\ref{eq:FFLO-sus})  due to quasiparticle scattering by using the symmetry $\tC_+(\bk\bk')=\tC_+(\bk'\bk)$ of coherence factors and the equivalence of summation over $\bk$~or $\bk'$ in the BZ. This leads to the form 
\be
\bl
\chi^{sc}_{0\bq}(\tbq,\omega)=
\frac{1}{2N}\sum_{\bk\si}
\tC^\bq_+(\bk\bk') 
\frac{f(E^\si_{\bk'\bq})-f(E^\si_{\bk\bq})}{\omega-(E^\si_{\bk'\bq}-E^\si_{\bk\bq})+i\eta}
.
\label{eq:qpscatt}
\el
\ee
Setting $b=0, q=0$  we have $\xi_\bk^s=\xi_\bk$ and $\xi^a_\bk=0$. Furthermore with  $\Delta_\bq^\bk= \Delta_0^\bk\equiv\Delta_\bk$ 
this leads to $E^\pm_{\bk\bq}=E_{\bk 0}\equiv E_\bk=[\xi_\bk^2+\Delta_\bk^2]^\fs$. This is the quasiparticle energy for the BCS case for {\it all} wave vectors since the pairing is stable for all values of~\bk ~and there is no more segmentation of the Fermi surface in paired and unpaired regions as in the FF case. This leads to a greatly simplified response function that only depends on the momentum transfer $\tbq$ and no longer on the FF pair vector $\bq$. We obtain from Eqs.~(\ref{eq:FFLO-sus},\ref{eq:qpscatt}):
\be
\bl
&
\chi_{0}(\tbq,\omega) =
\frac{1}{2N}\sum_\bk \Bigl\{
2\tC_+(\bk\bk')
\frac{f(E_{\bk'})-f(E_{\bk})}{\omega-(E_{\bk'}
\!-\!
E_{\bk})+i\eta}
+
\\
&
\tC_-(\bk\bk')
\Bigl[
\frac{1-f(E_{\bk'})-f(E_{\bk})}{\omega+(E_{\bk'}+E_{\bk})+i\eta}+
\frac{f(E_{\bk'})+f(E_{\bk})-1}{\omega-(E_{\bk'}+E_{\bk})+i\eta}
\Bigr]
\Bigr\}.
\label{eq:BCS-sus}
\el
\ee
The coherence factors now also simplify to
\be
\bl
\tC_\pm(\bk\bk')=\fs\Bigl[
1\pm \frac{\xi_{\bk}\xi_{\bk'}+\Delta_{\bk}\Delta_{\bk'}}
{E_{\bk}E_{\bk'}}\Bigr].
\label{eq:BCS-coh}
\el
\ee
This expression agrees with the result in Refs.~\cite{bulut:96,norman:00,michal:11}. 
For zero temperature and positive frequency only the last term in Eq.~(\ref{eq:BCS-coh}) contributes and the response function reduces to 
\be
\chi_{0}(\tbq,\omega) =
\frac{1}{2N}\sum_\bk\tC_-(\bk\bk')\frac{\Theta(-E_{\bk'})-\Theta(E_{\bk})}{\omega-(E_{\bk'}+E_{\bk})+i\eta}.
\ee
In the case of sign-changing unconventional gap function with $\Delta_{\bk+\bQ}=-\Delta_\bq$ where e.g. $\bQ=(\pi,\pi)$ for the d-wave gap function the coherence factor $\tC_-(\bk,\bk+\bQ)\simeq 1$ close to the gap threshold where $\xi_\bk=-\xi_{\bk+\bQ}\simeq 0$ (half filling) and $\min_{\bk\in FS}(|\Delta_\bk|+|\Delta_{\bk'}|)\approx 2\Delta_d$. On the other hand for the s-wave case with $\Delta_{\bk+\bQ}=\Delta_\bq=\Delta_0$ the coherence factor  $\tC_-(\bk,\bk+\bQ)\simeq 0$ is vanishingly small. This results in a step-like increase of ${\rm Im}\chi_0(\bQ,\omega)$ for $\omega >2\Delta_d$ in the d-wave case and only gradual increase for for $\omega >2\Delta_0$  for the s-wave gap. This is associated with a peak or no peak in the real part in both cases, respectively. Therefore a spin resonance in the collective RPA susceptibility at $\omega_r(\bQ)<2\Delta_d$ develops according to Eq.~(\ref{eq:resonance}) for the d-wave gap but not for the s-wave case \cite{thalmeier:16} (see Fig.~\ref{fig:susom}(a,b)).



\bibliography{References}

\end{document}